    \documentclass[pre,twocolumn,eqsecnum,showpacs]{revtex4}
\usepackage{amsmath}
    \usepackage{dcolumn}
    \usepackage{bm}
    \usepackage{graphicx}
    \usepackage{dcolumn}
\newcommand\beq{\begin{equation}}
\newcommand\eeq{\end{equation}}
\newcommand\beqa{\begin{eqnarray}}
\newcommand\eeqa{\end{eqnarray}}
\newcommand{\nn}{\nonumber\\}

        \newcommand\kk[1]{{{{{#1}}}}}

\begin{document}
\title{Modified Sonine approximation for the Navier--Stokes transport coefficients of a granular gas}

\author{Vicente Garz\'o}
\email{vicenteg@unex.es}
\homepage{http://www.unex.es/eweb/fisteor/vicente/}
\author{Andr\'es Santos}
\email{andres@unex.es}
\homepage{http://www.unex.es/eweb/fisteor/andres/}
\affiliation{Departamento de F\'{\i}sica, Universidad de
Extremadura, E-06071 Badajoz, Spain}
\author{Jos\'e Mar\'{\i}a Montanero}
\email{jmm@unex.es} \affiliation{Departamento de Electr\'onica e
Ingenier\'{\i}a Electromec\'anica, Universidad de Extremadura,
E-06071 Badajoz, Spain}

\begin{abstract}
Motivated by the disagreement found at high dissipation between
simulation data for the heat flux transport coefficients and the
expressions derived from the Boltzmann equation by the standard
first Sonine approximation [Brey et al., Phys. Rev. E \textbf{70},
051301 (2004); J. Phys.: Condens. Matter \textbf{17}, S2489 (2005)],
 we \kk{implement} in this paper a modified version  of the first Sonine
approximation in which the Maxwell--Boltzmann weight function is
replaced by  the homogeneous cooling state distribution.  The
structure of the transport coefficients is common in both
approximations,  the distinction  appearing in the coefficient of
the fourth cumulant $a_2$. Comparison with computer simulations
shows that  the modified approximation significantly improves the
estimates for the heat flux transport coefficients at strong
dissipation.  In addition, the slight discrepancies between
simulation and the standard first Sonine estimates for the shear
viscosity  and the self-diffusion coefficient are also partially
corrected by the modified approximation. Finally, the extension of
the modified first Sonine approximation to the transport
coefficients of the Enskog kinetic theory  is presented.\\


\end{abstract}


\draft
\date{\today}
\maketitle

\section{Introduction}
\label{sec1}

The usefulness of kinetic theory tools to describe the dynamical
properties of granular fluids has been widely recognized
\cite{C90,BP04}. The primary difference from normal fluids lies in
the inelastic character of collisions, what introduces features not
present in ordinary matter, such as the absence of equilibrium
states, the spontaneous formation of clusters, and the development
of high-energy tails, among others. The prototype model of a
granular gas is a system composed by a large number of smooth hard
spheres colliding inelastically with a constant coefficient of
normal restitution $0<\alpha\leq 1$. In the dilute limit, the
Boltzmann kinetic equation, suitably modified to incorporate
inelasticity, provides a convenient framework to investigate some of
the most relevant  properties of granular gases.

One of the main applications of the inelastic Boltzmann equation is
the derivation of the constitutive equations for the stress tensor
and the heat flux in a hydrodynamic description. These constitutive
equations define the relevant Navier--Stokes (NS) transport
coefficients and are determined by means of the Chapman--Enskog (CE)
method to solve the Boltzmann equation for sufficiently long space
and time scales. While in the elastic case the velocity distribution
function $f$ is expanded about the local equilibrium distribution
function $f_M$, the reference state for granular gases is the local
version $f^{(0)}$ of the so-called homogeneous cooling state (HCS).
The first-order stage of the CE expansion allows one to express the
transport coefficients  in terms of the solutions of linear integral
equations \cite{BDKS98,GD99,BC01,GD02,L05}. In these equations the
distribution $f^{(0)}$ appears explicitly in the inhomogeneous terms
and implicitly through the linearized Boltzmann collision operator.
As happens in the elastic case, the solutions of the integral
equations  are not  known exactly, so that approximations must be
introduced in order to  get  the transport coefficients as explicit
nonlinear functions of $\alpha$. The standard method consists of
approximating the solutions  by the Maxwell--Boltzmann distribution
$f_M$ times truncated Sonine polynomial expansions. The simplest
possibility is the first Sonine approximation, where only the lowest
Sonine polynomial is retained. The resulting expressions for the
effective collision frequencies associated with the transport
coefficients have an explicit dependence on $\alpha$ (due to the
collision rules) as well as an implicit one through  a linear
dependence on the fourth cumulant $a_2$ of $f^{(0)}$.

The reliability of the standard first Sonine approximation has been
tested in the last few years
\cite{BRMC99,BRMCGR00,LBD02,GM02,MG03,GM04,BRM04,BRMMG05,MSG05,MSG06}
by comparison with computer simulations of the Boltzmann equation by
means of the direct simulation Monte Carlo (DSMC) method
\cite{DSMC}. The comparisons show that the shear viscosity $\eta$
and the self-diffusion coefficient $D$ are accurately estimated by
the first Sonine approximation, even for strong dissipation. The two
transport coefficients $\kappa$ and $\mu$ characterizing the heat
flux are well described by the first Sonine approximation for
moderate and small inelasticity (say $\alpha\gtrsim 0.7$). However,
recent studies \cite{BRM04,BRMMG05} show that the first Sonine
approximation dramatically overestimates $\kappa$ and $\mu$ for high
dissipation ($\alpha\lesssim 0.7$).

Although the range $\alpha\gtrsim 0.7$ encompasses the region of
practical and experimental interest, especially if one considers the
inherent coupling between gradients and dissipation in steady states
\cite{BC98,SGD04}, it is important from a fundamental point of view
to understand the origin of those discrepancies for $\kappa$ and
$\mu$ and propose alternative approximations. Since in the
simulations carried out in Refs.\ \cite{BRM04,BRMMG05} the transport
coefficients were obtained from two-time correlation functions by
means of Green--Kubo (GK) relations, the discrepancies might be due
to velocity correlation effects outside the domain of the Boltzmann
equation. However, as discussed in Ref.\ \cite{MSG06}, the
disagreement seems to be directly related to the failure of the
Sonine expansion truncated after the first term to capture the
velocity dependence of the NS distribution function. One possibility
of improving the approximation could be to consider higher order
terms in the Sonine expansion \cite{GM04}. However, the involved
algebra would be rather intricate and it is not obvious that the
improvement would be significant.

The aim of this paper is to \kk{implement} an alternative route to
the standard first Sonine approximation. The idea is based on the
assumption that the isotropic part of the NS velocity distribution
$f^{(1)}$ is mainly governed by the HCS distribution $f^{(0)}$
rather than by the Maxwellian distribution $f_M$ \kk{\cite{L05}}.
More specifically, our modified first Sonine approximation has the
same form as the standard one, except that the weight function $f_M$
is replaced by $f^{(0)}$. As a consequence, the effective collision
frequencies derived from the modified approximation have the same
structure as those derived from the standard one, except that the
respective coefficients of $a_2$ differ markedly in both
approximations. Since the high-velocity population in $f^{(0)}$ is
larger than in $f_M$, it is reasonably expected that the former
captures better the influence of the high-velocity tail of $f^{(1)}$
on the NS transport coefficients, especially those related to the
heat flux. In fact, the results show that, in the region
$\alpha\lesssim 0.7$, the modified values for the collision
frequencies are larger than their standard counterparts, so that the
corresponding transport coefficients are smaller in the modified
first Sonine approximation than in the standard one. As will be
shown later, the modified estimates for  the transport coefficients
$\kappa$ and $\mu$ compare quite well with available computer
simulations, even for extreme dissipation, in contrast to what
happens with the standard estimates. For the remaining transport
coefficients $\eta$ and $D$, which are already well described by the
standard approximation, the modified approximation provides even
better values.

The plan of the paper is as follows. The expressions of the NS
transport coefficients obtained by the application of the standard
first Sonine approximation  are recalled in  Sec.\ \ref{sec3}.  Our
modified first Sonine approximation is described and discussed in
Sec.\ \ref{sec4}. Next, the two Sonine approximations are compared
in Sec\ \ref{sec5} with available and new simulation data for the
transport coefficients $\kappa$, $\mu$, $\eta$, and $D$, both for
two- and three- dimensional systems. The extension to the transport
coefficients provided by the Enskog theory is presented in Appendix
\ref{appB}. The paper is closed with some concluding remarks in
Sec.\ \ref{sec6}.

\section{Standard first Sonine approximation\label{sec3}}

We consider a granular gas of smooth, inelastic hard spheres (in $d$
dimensions) of mass $m$, diameter $\sigma$, and coefficient of
restitution $\alpha$. In  the low-density regime, the one-particle
velocity distribution function $f({\bf r},{\bf v},t)$ obeys the
(inelastic) Boltzmann equation \cite{GS95,BDS97}.
 Under the conditions of weak hydrodynamic gradients, the  CE method
\cite{CC70} provides a solution of the Boltzmann equation based on
an expansion $f=f^{(0)} +f^{(1)}+\cdots$, where  $f^{(0)}$ is the
{\em local} version of the HCS \cite{GS95,vNE98}. The first-order
distribution $f^{(1)}$ has the form \cite{BDKS98,BC01}
\begin{equation}
\label{2.16.1}
f^{(1)}({\bf V})={\bm{\mathcal{A}}}({\bf V})\cdot \nabla \ln
T+{\bm{\mathcal{B}}}({\bf V})\cdot \nabla \ln n+{\cal C}_{ij}({\bf
V}) \nabla_iu_j,
\end{equation}
where $n$, $T$, and $\mathbf{u}$ are the number density, granular
temperature, and flow velocity, respectively, and
$\mathbf{V}=\mathbf{v}-\mathbf{u}$ is the peculiar velocity. The
functions ${\bm{\mathcal{A}}}({\bf V})$, ${\bm{\mathcal{B}}}({\bf
V})$, and ${\cal C}_{ij}({\bf V})$ are the solutions of a set of
linear integral equations. While  ${\bm{\mathcal{A}}}({\bf V})$ and
${\cal C}_{ij}({\bf V})$  obey autonomous equations, the equation
for ${\bm{\mathcal{B}}}({\bf V})$ requires the knowledge of
${\bm{\mathcal{A}}}({\bf V})$. However, the combination
${\bm{\mathcal{A}}}'({\bf V})={\bm{\mathcal{A}}}({\bf
V})-\frac{1}{2}{\bm{\mathcal{B}}}({\bf V})$ satisfies a closed
equation \cite{MSG06}. From the solutions to those linear integral
equations,  the   shear viscosity $\eta$ (associated with the
pressure tensor), the thermal conductivity $\kappa$, and  a new
 transport coefficient $\mu$  (the two latter associated with the heat-flux) are formally given by \cite{BDKS98,BC01,MSG06}
\begin{equation}
\label{2.19}
\eta=\eta_0\frac{1}{\nu_{\eta}^*-\frac{1}{2}\zeta^{*}},
\end{equation}
\begin{equation}
\label{2.20}
\kappa=\kappa_0 \frac{d-1}{d}\frac{1+2a_2}{\nu_{\kappa}^*-2
\zeta^*},
\end{equation}
\beq
\mu=\frac{2T}{n}(\kappa-\kappa'), \quad \kappa'=\kappa_0
\frac{d-1}{d}\frac{1+\frac{3}{2}a_2}{\nu_{\kappa'}^*-\frac{3}{2}
\zeta^*},
\label{2.21}
\eeq
where, $\eta_0$ and $\kappa_0$ are the elastic values (in the first
Sonine approximation) of the shear viscosity and thermal
conductivity \cite{CC70}, respectively [cf.\ Eq.\ \eqref{2.22}]. In
addition, $\zeta^*=\zeta^{(0)}/\nu_0$ is the reduced cooling rate of
the HCS, where $\nu_0=nT/\eta_0$ is an effective collision frequency
[cf.\ Eq.\ \eqref{x3.14b}], and $a_2$ is the fourth velocity
cumulant of $f^{(0)}$. Furthermore, in Eqs.\
(\ref{2.19})--(\ref{2.21}) the (reduced) effective collision
frequencies are
\begin{equation}
\label{2.25}
\nu_\eta^*=\frac{\int d{\bf v} D_{ij}({\bf V}){\cal L}{\cal C}_{ij}({\bf V})}
{\nu_0\int d{\bf v}D_{ij}({\bf V}){\cal C}_{ij}({\bf V})},
\end{equation}
\begin{equation}
\label{2.26}
\nu_\kappa^*=\frac{\int d{\bf v} {\bf S}({\bf V})\cdot{\cal L}
{\bm{\mathcal{A}}}({\bf V})} {\nu_0\int d{\bf v}{\bf S}({\bf
V})\cdot{\bm{\mathcal{A}}}({\bf V})}, \quad
\nu_{\kappa'}^*=\frac{\int d{\bf v} {\bf S}({\bf V})\cdot{\cal
L}{\bm{\mathcal{A}}}'({\bf V})} {\nu_0\int d{\bf v}{\bf S}({\bf
V})\cdot{\bm{\mathcal{A}}}'({\bf V})},
\end{equation}
where ${\cal L}$ is the linearized Boltzmann collision operator and
we have introduced the polynomials
\beq
D_{ij}(\mathbf{V})=m\left({V}_i{V}_j-\frac{1}{d}V^2\delta_{ij}\right),
\label{n3bis}
\eeq
\beq
\mathbf{S}(\mathbf{V})=\left(\frac{m}{2}
V^{2}-\frac{d+2}{2}T\right){\bf V}.
\label{n4bis}
\eeq

Another relevant transport coefficient for a single gas is the
self-diffusion coefficient $D$, which measures the diffusion of
tagged particles in a fluid of mechanically equivalent particles in
the HCS. Application of the CE method  leads to $f_s^{(0)}=(n_s/n)
f^{(0)}$ and
\beq
f^{(1)}_s(\mathbf{V})=\bm{\mathcal{D}}(\mathbf{V})\cdot\nabla n_s,
\label{f1s}
\eeq
where $n_s$ and $f_s$ are the number density and velocity
distribution function of the tagged particles, respectively. As
before, the function $\bm{\mathcal{D}}(\mathbf{V})$ is the solution
of a linear integral equation. The CE expression for the
self-diffusion coefficient $D$ is \cite{BRMCGR00,GM04}
\begin{equation}
\label{2.29.3}
D=D_0\frac{d+2}{2d}\frac{1}{\nu_D^*-\frac{1}{2}\zeta^*},
\end{equation}
where $D_0$ is the elastic value of the self-diffusion coefficient
(in the first Sonine approximation) [cf.\ Eq.\ \eqref{2.29.3bis}]
and
\begin{equation}
\label{2.29.5}
\nu_D^*=\frac{\int d{\bf v}\, {\bf V}\cdot
\mathcal{L}_s\bm{\mathcal{D}}({\bf V})} {\nu_0\int d{\bf v}\,{\bf
V}\cdot {\bm {\mathcal{D}}}({\bf V})}
\end{equation}
is the (reduced) collision frequency associated with the
self-diffusion coefficient, $\mathcal{L}_s$ being the
Boltzmann--Lorentz operator.

So far,  the expressions for the NS transport coefficients are
formally exact, but their $\alpha$-dependence through the quantities
$\zeta^*$, $a_2$, $\nu_{\eta}^*$, $\nu_{\kappa}^*$,
$\nu_{\kappa'}^*$, and $\nu_D^*$ is not explicitly known. The two
first quantities ($\zeta^*$ and $a_2$) depend on the HCS
distribution  and are very accurately estimated  by
\cite{vNE98,MS00,CDPT03}
\beq
\zeta^*=\frac{d+2}{4d}(1-\alpha^2)\left(1+\frac{3}{16}a_2\right),
\label{x2.29}
\eeq
\beq
a_2= \frac{16(1-\alpha)(1-2\alpha^2)}{25+24d-\alpha (57-
8d)-2(1-\alpha)\alpha^2}.
\label{2.12}
\eeq
However, the determination of the collision frequencies
$\nu_{\kappa}^*$, $\nu_{\kappa'}^*$,  $\nu_{\eta}^*$, and $\nu_D^*$
is much more complicated since it requires the knowledge of
$\bm{\mathcal{A}}(\mathbf{V})$, $\bm{\mathcal{A}}'(\mathbf{V})$,
$\mathcal{C}_{ij}(\mathbf{V})$, and $\bm{\mathcal{D}}(\mathbf{V})$,
respectively. These functions are the solutions of linear integral
equations in which the HCS distribution $f^{(0)}$ appears explicitly
in the inhomogeneous terms and also implicitly through the
linearized Boltzmann operator. {}From that point of view, those
collision frequencies are \textit{functionals} of  $f^{(0)}$. In
order to get  explicit expressions for the dependence of the
transport coefficients on $\alpha$ one has to resort to some
approximations. As in the elastic case, the simplest approximation
consists of truncating the Sonine polynomial expansions of
${\bm{\mathcal{A}}}({\bf V})$, ${\bm {\mathcal{A}}}'({\bf V})$,
${\cal C}_{ij}({\bf V})$, and ${\bm{\mathcal{D}}}({\bf V})$ after
the first term. More explicitly, the first Sonine approximation is
\begin{equation}
\label{2.30x}
\left(
\begin{array}{c}
{\bm{\mathcal{A}}}({\bf V}) \\
{\bm{\mathcal{A}}}'({\bf V}) \\
{\cal C}_{ij}({\bf V})\\
{\bm {\mathcal{D}}}({\bf V})
\end{array}
\right) \rightarrow f_M(\mathbf{V})\left(
\begin{array}{c}
c_{\kappa}{\bf S}({\bf V}) \\
c_{\kappa'}{\bf S}({\bf V}) \\
c_{\eta}D_{ij}({\bf V})\\
c_D {\bf V}
\end{array}
\right) ,
\end{equation}
where the Maxwellian
\beq
f_{M}(\mathbf{V})=n\left(\frac{m}{2\pi T}\right)^{d/2} e^{-mV^2/2T}
\label{fM}
\eeq
is the weight factor in the scalar product with respect to which the
orthogonal polynomials are defined. The coefficients $c_\kappa$,
$c_{\kappa'}$, $c_\eta$, and $c_D$ are directly related to the
transport coefficients  by
\beq
\label{n2.32}
\kappa=-\frac{d+2}{2}\frac{nT^2}{m}c_\kappa,
\eeq
\beq
\label{n2.33}
\kappa'=-\frac{d+2}{2}\frac{nT^2}{m}c_{\kappa'},
\eeq
\beq
\label{n2.34}
\eta=-nT^2 c_\eta,
\eeq
\beq
\label{n2.34.1}
D=-\frac{n T}{m}c_D.
\eeq
Note that the approximations (\ref{2.30x}) imply that
$\bm{\mathcal{A}}'(\mathbf{V})\propto \bm{\mathcal{A}}(\mathbf{V})$,
what leads to $\nu_{\kappa'}^*=\nu_{\kappa}^*$.

 Inserting the
approximations (\ref{2.30x}) into Eqs.\ (\ref{2.25}), (\ref{2.26}),
and (\ref{2.29.5}), and taking the Sonine approximation for
$f^{(0)}$, one can evaluate explicitly the (reduced) collision
frequencies (\ref{2.25}), (\ref{2.26}), and  (\ref{2.29.5}). The
results are \cite{BDKS98,BRMCGR00,BC01}
\beqa
\label{2.36x}
\nu_\kappa^*=\nu_{\kappa'}^*&=&
\frac{1+\alpha}{d}\left[\frac{d-1}{2}+\frac{3}{16}(d+8)(1-\alpha)\right.\nn
&&\left.+\frac{4+5d-3(4-d)\alpha}{512}a_2\right],
\eeqa
\begin{equation}
\label{2.35x}
\nu_\eta^*=\frac{3}{4d}\left(1-\alpha+\frac{2}{3}d\right)(1+\alpha)\left(1
-\frac{1}{32}a_2\right),
\end{equation}
\begin{equation}
\label{2.37x}
\nu_D^*= \frac{d+2}{4d}(1+\alpha)\left(1-\frac{1}{32}a_2\right).
\end{equation}
Therefore, the NS transport coefficients in the standard first
Sonine approximation are given by Eqs.\ (\ref{2.19})--(\ref{2.21})
and (\ref{2.29.3}) with the collision frequencies given by Eqs.\
(\ref{2.36x})--(\ref{2.37x}). In addition, the cooling rate
$\zeta^*$ and the fourth cumulant $a_2$ are given by Eqs.\
(\ref{x2.29}) and (\ref{2.12}), respectively.

As said in the Introduction, while the results obtained in the first
Sonine approximation for the shear viscosity
\cite{BRMC99,BRM04,BRMMG05,MSG05} and the diffusion coefficient
\cite{BRMCGR00,LBD02,GM04} compare quite well with computer
simulations over a wide range of inelasticities, the coefficients
$\kappa$ and $\mu$ associated with  the heat flux show important
discrepancies with simulation data for strong inelasticity
\cite{BRM04,BRMMG05,MSG06}.

\section{Modified first Sonine approximation}
\label{sec4}
One of the possible sources of discrepancy between the standard
first Sonine approximation for the transport coefficients associated
with the heat flux and computer simulations could be due to the
existence of non-Gaussian features. Although the Maxwellian
distribution $f_M$ is a good approximation to $f^{(0)}$ in the
region of thermal velocities relevant to low-order moments
(hydrodynamic quantities), quantitative discrepancies between both
distributions are expected to be important  in the case of higher
velocity moments, such as the heat flux. The departure of $f^{(0)}$
from $f_M$ is partially accounted for by $a_2$. However, in the
approximation (\ref{2.30x}) the behavior of $f^{(1)}$ is assumed to
be essentially dominated by the Maxwellian distribution $f_M$.
{}From that point of view, one might say that a certain mismatch
exists in the standard first Sonine approximation applied to
inelastic gases. This could be fixed by incorporating more terms in
the Sonine polynomial expansion \cite{GM04}, but this would be at
the expense of significantly increasing the technical difficulties
of the method.

Here we follow an alternative route, \kk{similar to the one
discussed in Ref.\ \cite{L05}}. Specifically, we keep the structure
of (\ref{2.30x}), except that the distribution $f^{(0)}$ is chosen
instead of the simple Maxwellian form $f_{M}$ as the convenient
weight function. According to these arguments,  we take the
approximations
\begin{equation}
\label{2.30}
\left(
\begin{array}{c}
{\bm{\mathcal{A}}}({\bf V}) \\
{\bm{\mathcal{A}}}'({\bf V}) \\
{\cal C}_{ij}({\bf V})\\
{\bm {\mathcal{D}}}({\bf V})
\end{array}
\right) \rightarrow f^{(0)}(\mathbf{V})\left(
\begin{array}{c}
\overline{c}_{\kappa}\overline{\bf S}({\bf V}) \\
\overline{c}_{\kappa'}\overline{\bf S}({\bf V}) \\
\overline{c}_{\eta}\overline{D}_{ij}({\bf V})\\
\overline{c}_D {\bf V}
\end{array}
\right) ,
\end{equation}
where  $\overline{\bf S}({\bf V})$ and $\overline{D}_{ij}({\bf V})$
have the same polynomial structure as ${\bf S}({\bf V})$ and
${D}_{ij}({\bf V})$, respectively, but must be chosen to preserve
the \kk{solubility  conditions \cite{CC70,GS03}}. A simple
calculation yields
\begin{equation}
\label{2.31a}
\overline{D}_{ij}({\bf V})=D_{ij}({\bf V}),
\end{equation}
\beq
\overline{\bf S}({\bf V})={\bf S}({\bf V})-\frac{d+2}{2}a_2T{\bf V}.
\label{2.31b}
\eeq
As before, $\bm{\mathcal{A}}'(\mathbf{V})\propto
\bm{\mathcal{A}}(\mathbf{V})$, so that
$\nu_{\kappa'}^*=\nu_{\kappa}^*$ in the modified first Sonine
approximation also. {} \kk{The coefficients $\overline{c}_{\kappa}$,
$\overline{c}_{\kappa'}$, $\overline{c}_{\eta}$, and
$\overline{c}_{D}$ are related to the transport coefficients by}
\beq
\label{2.32}
\kappa=-\frac{d+2}{2}\frac{nT^2}{m}\left(1+\frac{d+8}{2}a_2-\frac{d+2}{2}a_2^2-\frac{d+4}{2}{a_3}\right)\overline{c}_{\kappa},
\eeq
\beq
\label{2.33}
\kappa'=\kappa\frac{\overline{c}_{\kappa'}}{\overline{c}_{\kappa}},
\eeq
\beq
\label{2.34}
\eta=-nT^2 \left(1+a_2\right)\overline{c}_{\eta},
\eeq
\beq
\label{2.34.1}
D=-\frac{n T}{m}\overline{c}_D.
\eeq
\kk{In Eq.\ (\ref{2.32}), $a_3$ is the sixth cumulant of $f^{(0)}$
\cite{vNE98}.} Note that in Eqs.\ (\ref{2.31a})--(\ref{2.34.1}) no
explicit form for $f^{(0)}$ has been \kk{needed to be} assumed.

In the modified first Sonine approximation, the collision
frequencies are obtained from Eqs.\ (\ref{2.25}), (\ref{2.26}), and
(\ref{2.29.5}) by inserting the approximations (\ref{2.30}), and
neglecting $a_3$ and nonlinear terms in $a_2$. After  lengthy
algebra \cite{note}, one gets
\beqa
\label{2.36}
\nu_\kappa^*=\nu_{\kappa'}^*&=&
\frac{1+\alpha}{d}\left[\frac{d-1}{2}+\frac{3}{16}(d+8)
(1-\alpha)\right.\nn
&&\left.+\frac{296+217d-3(160+11d)\alpha}{256}a_2\right],
\eeqa
\begin{equation}
\label{2.35}
\nu_\eta^*=\frac{3}{4d}\left(1-\alpha+\frac{2}{3}d\right)(1+\alpha)
\left(1+\frac{7}{16}a_2\right),
\end{equation}
\begin{equation}
\label{2.37}
\nu_D^*= \frac{d+2}{4d}(1+\alpha)\left(1+\frac{3}{16}a_2\right).
\end{equation}
Thus, the NS transport coefficients in the modified first Sonine
approximation are given by Eqs.\ (\ref{2.19})--(\ref{2.21}) and
(\ref{2.29.3}), with the collision frequencies given by Eqs.
(\ref{2.36})--(\ref{2.37}).

Comparison between the standard approximations, Eqs.\
(\ref{2.36x})--(\ref{2.37x}), and the modified ones, Eqs.\
(\ref{2.36})--(\ref{2.37}), shows that they only differ in the
coefficient of the term linear in $a_2$. In the standard
approximation, the dependence of the collision frequencies on $a_2$
only arises from the presence of the HCS distribution $f^{(0)}$ in
the linear operators $\mathcal{L}$ and $\mathcal{L}_s$. On the other
hand, in the modified approximation there exist additional
contributions arising from the weight factor $f^{(0)}$ in Eq.\
(\ref{2.30}) and, in the case of $\nu_{\kappa}^*$, also from the
modified Sonine polynomial $\overline{\mathbf{S}}(\mathbf{V})$.
These additional contributions give rise to a renormalization of the
coefficients of $a_2$, which change dramatically with respect to
their values in the standard approximation. More specifically, the
coefficient of $a_2$ in Eq.\ (\ref{2.36}) is at least 46 times
larger than the coefficient in Eq.\ (\ref{2.36x}), both being
positive. In the cases of $\nu_{\eta}^*$ and $\nu_{D}^*$, the
coefficients are negative in the standard first Sonine
approximation, while they are positive in the modified Sonine
approximation. Moreover, the magnitudes of the coefficients in the
latter approximation are 14 and 6 times larger, respectively, than
in the former one. These discrepancies are not significant as long
as the magnitude of $a_2$ is relatively small. This is what happens
for $\alpha\gtrsim 0.7$. However, for larger inelasticity, the
fourth cumulant $a_2$ is not negligible
\kk{\cite{MS00,CDPT03,BCRM99,BP06}}. Since $a_2>0$ for
$\alpha\lesssim 0.7$, then the standard estimates for the collision
frequencies are smaller than their modified counterparts.
Consequently, the associated transport coefficients are larger in
the standard approximation than in the modified one. The fact that
these effective collision frequencies associated with the transport
coefficients are larger in the modified approximation than in the
standard one is possibly due to the overpopulation of $f^{(0)}$ with
respect to $f_M$ for high velocities. Since the collision rate for
hard spheres increases with velocity,  this overpopulation yields a
more efficient average collisional transfer of momentum and energy.

\kk{In principle, Eq.\ (\ref{2.30}) can be seen as the first-order
approximation in a polynomial expansion. For instance, in the case
of ${\bm{\mathcal{A}}}({\bf V})$ one can write
\beq
{\bm{\mathcal{A}}}({\bf
V})=f^{(0)}(\mathbf{V})\mathbf{V}\sum_{k=1}^\infty c_k
\overline{L}_k^{(d/2)}(c^2),
\label{4.x1}
\eeq
where $\{\overline{L}_k^{(d/2)}(x)\}$ is a set of orthogonal
polynomials with respect to an inner product involving $f^{(0)}$. If
$f^{(0)}$ is replaced by $f_M$, then the polynomials
$\overline{L}_k^{(d/2)}(x)$ become the generalized Laguerre
polynomials ${L}_k^{(d/2)}(x)$. The expansion (\ref{4.x1}) differs
from the one considered in Ref.\ \cite{L05} in the use of the
modified polynomials $\overline{L}_k^{(d/2)}(x)$ instead of the
conventional polynomials ${L}_k^{(d/2)}(x)$, which do not constitute
an orthogonal set in this case. The recursive procedure to get the
polynomials $\overline{L}_k^{(d/2)}(x)$ in terms of the cumulants of
$f^{(0)}$ is briefly described in Appendix \ref{appC}. }

\section{Comparison with computer simulations\label{sec5}}
In this Section we compare the theoretical expressions  for the
transport coefficients obtained from the standard and modified first
Sonine approximations with available and new computer simulations.
\subsection{Heat flux}
The NS transport coefficients associated with the heat flux are
$\kappa$ and $\mu$.  These transport coefficients have been measured
from the GK relations \cite{DB02} by means of the DSMC method
\cite{DSMC}, both for two- \cite{BRM04} and three-dimensional
\cite{BRMMG05} systems. In addition, the coefficient
$\kappa'=\kappa-n\mu/2T$ has been measured in DSMC simulations by an
alternative method based on the application of an external force  in
the three-dimensional case \cite{MSG06}.

\begin{figure}[htb]
\includegraphics[width=\columnwidth]{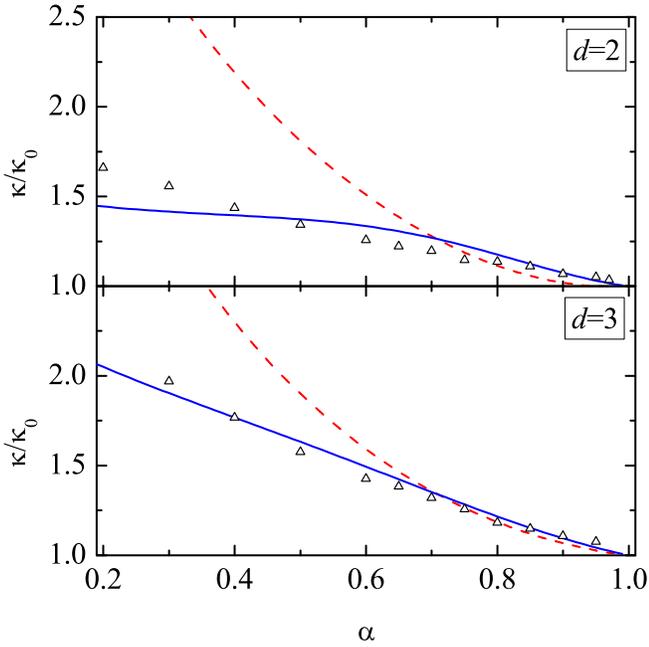}
\caption{(Color online) Plot of the  (reduced) thermal conductivity
$\kappa/\kappa_0$  as a function of $\alpha$ for hard disks (top
panel)  and hard spheres (bottom panel). The dashed and solid lines
represent the standard and modified first Sonine approximations,
respectively. The symbols are DSMC results obtained from the GK
relations \protect\cite{BRM04,BRMMG05}.
\label{fig1}}
\end{figure}

\begin{figure}[htb]
\includegraphics[width=\columnwidth]{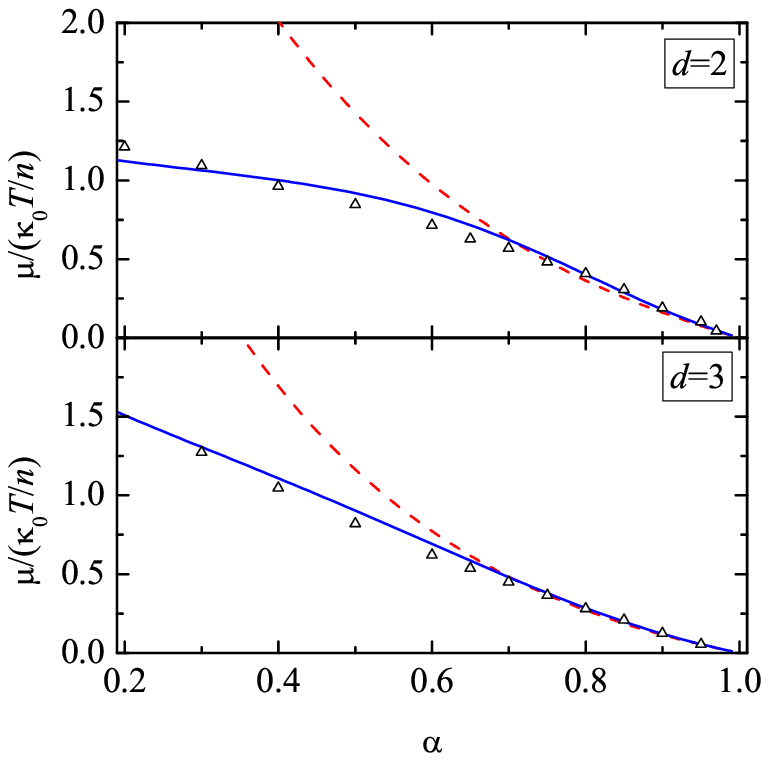}
\caption{(Color online) Plot of the  (reduced) transport coefficient
 $\mu/(\kappa_0T/n)$  as a function of $\alpha$
for hard disks (top panel)  and hard spheres (bottom panel). The
dashed and solid lines represent the standard and modified first
Sonine approximations, respectively. The symbols are DSMC results
obtained from the GK relations \protect\cite{BRM04,BRMMG05}.
\label{fig2}}
\end{figure}

Figures \ref{fig1} and \ref{fig2} show the $\alpha$-dependence of
the reduced transport coefficients $\kappa/\kappa_0$ and
$\mu/(T\kappa_0/n)$, respectively. It is apparent that the standard
first Sonine approximation significantly overestimates both
transport coefficients for strong inelasticity. On the other hand,
the modified approximation compares  well with computer simulations,
even for low values of $\alpha$, especially in the three-dimensional
case. This reflects the fact that the modified approximation is more
accurate than the standard one in describing the effective collision
frequencies for heat transport.

\begin{figure}[htb]
\includegraphics[width=\columnwidth]{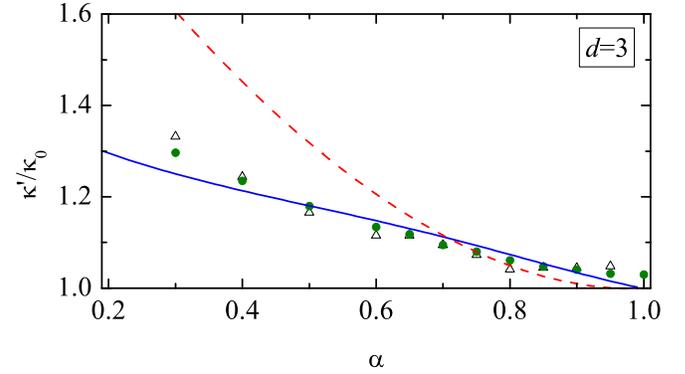}
\caption{(Color online) Plot of the  (reduced)  modified thermal
conductivity $\kappa'/\kappa_0$  as a function of $\alpha$ for hard
spheres. The dashed and solid lines represent the standard and
modified first Sonine approximations, respectively. The symbols
 are DSMC results obtained from the GK relations
 \protect\cite{BRMMG05} (triangles) and from the application of an external
force \protect\cite{MSG06} (circles).
\label{fig3}}
\end{figure}

Since both $\kappa$ and $\mu$ are overestimated by the standard
approximation, it could happen that, by a cancelation of errors, the
transport coefficient $\kappa'=\kappa-n\mu/2T$ might be well
captured by that approximation.  However, this is not the case. The
comparison between the computer simulations for $\kappa'$ obtained
from the two alternative methods of Refs.\ \cite{BRMMG05} and
\cite{MSG06} and the two theoretical approaches is shown in Fig.
\ref{fig3}. As in the cases of $\kappa$ and $\mu$, the modified
Sonine approximation for $\kappa'$ agrees quite well with the
simulation results.

The good agreement found in Figs.\ \ref{fig1}--\ref{fig3} between
the modified first Sonine approximation and the simulation data for
the heat flux transport coefficients suggests that the NS
distribution functions $\bm{\mathcal{A}}(\mathbf{V})$ and
$\bm{\mathcal{A}}'(\mathbf{V})$ are well represented by the forms
(\ref{2.30}). To test this expectation, we compare now the standard
and  modified Sonine approximations for
$\bm{\mathcal{A}}'(\mathbf{V})$ with simulation data presented in
Ref.\ \cite{MSG06} for the three-dimensional case. By symmetry
arguments, the function $\bm{\mathcal{A}}'(\mathbf{V})$ can be
written as
\beq
\bm{\mathcal{A}}'(\mathbf{V})=\lambda
v_0^{-1}f_M(\mathbf{V})\Phi(c^2)\mathbf{V},
\label{5.1}
\eeq
where $\lambda=1/\sqrt{2}\pi n\sigma^2$ is the mean free path,
$v_0=\sqrt{2T/m}$ is the thermal speed, and $\Phi(c^2)$ is a
dimensionless isotropic function of the scaled velocity $
\mathbf{c}={\mathbf{V}}/{v_0}$.
 All the information contained in $\Phi(c^2)$ is retained by the
marginal distribution \cite{MSG06}
\beq
\varphi(c_x^2)=\pi^{-1}\int_{-\infty}^\infty dc_y
\int_{-\infty}^\infty dc_z\, e^{-(c_y^2+c_z^2)}\Phi(c^2).
\label{5.3}
\eeq
According to the standard approximation (\ref{2.30x}),
\beq
\Phi(c^2)=\frac{4\kappa'}{5n\lambda
v_0}\left(\frac{5}{2}-c^2\right),
\label{5.3bis}
\eeq
so that
\beq
\varphi(c_x^2)=\frac{4\kappa'}{5n\lambda
v_0}\left(\frac{3}{2}-c_x^2\right).
\label{5.4}
\eeq
In contrast, the modified approximation (\ref{2.30}) yields,
\beqa
\Phi(c^2)&=&\frac{4\kappa'}{5n\lambda
v_0}\left[1+\frac{a_2}{2}\left(c^4-5c^2+\frac{15}{4}\right)\right]\frac{\frac{5}{2}(1+a_2)-c^2}{1+\frac{11}{2}a_2}\nn
&=&\frac{4\kappa'}{5n\lambda v_0}\left[\frac{5}{2}-c^2+a_2\left(
\frac{105}{16}+\frac{21}{8}c^2\right.\right.\nn &&\left.\left.-
\frac{15}{4}c^4+\frac{1}{2}c^6\right) \right],
\label{5.5}
\eeqa
where in the last equality we have neglected nonlinear terms in
$a_2$. The corresponding marginal distribution is
\beq
\varphi(c_x^2)=\frac{4\kappa'}{5n\lambda
v_0}\left[\frac{3}{2}-c_x^2-a_2\left(
\frac{75}{16}-\frac{15}{8}c_x^2-
\frac{9}{4}c_x^4+\frac{1}{2}c_x^6\right) \right].
\label{5.6}
\eeq

\begin{figure}[htb]
\includegraphics[width=\columnwidth]{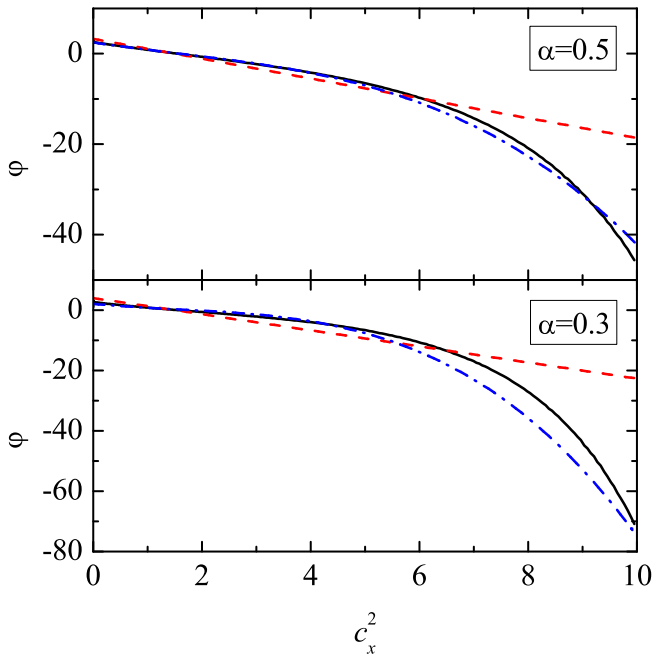}
\caption{(Color online)  Plot of the  marginal distribution function
$\varphi(c_x^2)$ obtained from DSMC (solid lines) for $\alpha=0.5$
(top panel) and $\alpha=0.3$ (bottom panel).  The  dashed and
dotted-dashed lines represent the standard and modified first Sonine
approximations, respectively.
\label{fig4}}
\end{figure}

The function $\varphi(c_x^2)$ is plotted in Fig.\ \ref{fig4} for
$\alpha=0.5$ and $\alpha=0.3$. We see that the modified first Sonine
distribution (\ref{5.6}) captures reasonably well the main features
of the true distribution, especially for $\alpha=0.5$. On the other
hand, the standard first Sonine distribution (\ref{5.4}) strongly
disagrees with the simulation data for $c_x^2\gtrsim 6$. This
velocity region has a significant influence on the evaluation of the
thermal conductivity at high inelasticity \cite{MSG06}.

\subsection{Pressure tensor}
\begin{figure}[htb]
\includegraphics[width=\columnwidth]{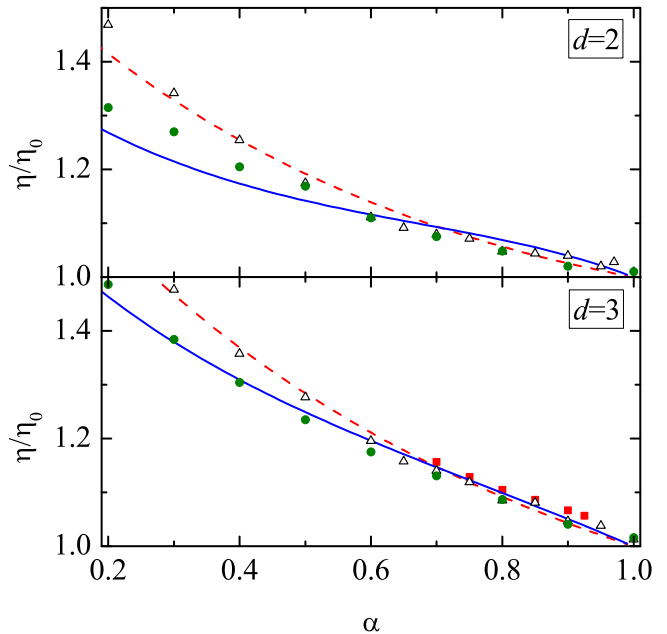}
\caption{(Color online) Plot of the  (reduced) shear viscosity
$\eta/\eta_0$ as a function of $\alpha$ for hard disks (top panel)
and hard spheres (bottom panel). The dashed and solid lines
represent the standard and modified first Sonine approximations,
respectively. The symbols  are DSMC results obtained from the decay
of a sinusoidal perturbation \protect\cite{BRMC99} (squares), from
the GK relations \protect\cite{BRM04,BRMMG05} (triangles), and from
the application of an external force \protect\cite{MSG05} (circles).
In the latter case, the data corresponding  to $d=3$ for $\alpha\leq
0.5$ and  to $d=2$ have been obtained in the present work.
\label{fig5}}
\end{figure}

Although the $\alpha$-dependence of the shear viscosity is well
described by the standard first Sonine approximation, it is
worthwhile comparing the modified first Sonine estimate against
computer simulations. To the best of our knowledge, the NS shear
viscosity has been measured in DSMC simulations by three alternative
methods: (i) by analyzing the time decay of a weak transverse shear
wave in the HCS \cite{BRMC99}; (ii) from the GK relation
\cite{BRM04,BRMMG05}; and (iii) by the application of a homogeneous
external force \cite{MSG05}. The simulation data obtained by these
methods and the two theoretical approximations are presented in
Fig.\ \ref{fig5}. The data obtained from the method (i) are
restricted to $d=3$ and $\alpha\geq 0.7$ \cite{BRMC99}, while the
ones from the method (ii) are available for $d=2$ \cite{BRM04} and
$d=3$ \cite{BRMMG05}. Regarding the method (iii), the data
corresponding to $d=3$ for $\alpha\geq 0.6$ were reported in Ref.\
\cite{MSG05}, while those corresponding  to $d=3$ for $\alpha\geq
0.5$ and to $d=2$ have been obtained in this work.  We observe that
up to $\alpha\simeq 0.6$ the simulation data are consistent among
themselves and also with both theories. However, for higher
inelasticities, there is a certain discrepancy (less than 10\%)
between the data reported in Refs.\ \cite{BRM04,BRMMG05} and those
presented here, the former being close to the standard estimates and
the latter being close to the modified estimates, especially in the
three-dimensional case. \kk{The small difference between our
simulation data and those of Refs.\ \cite{BRM04,BRMMG05} might be
due to the influence of  velocity correlations in the correlation
function involved in the GK expression of the shear viscosity. These
velocity correlations are larger than in the case of the heat-flux
transport coefficients \cite{BRM04}.}

 In conclusion, while the standard first
Sonine approximation does quite good a job for the shear viscosity,
it is fair to say that the modified approximation is still better,
especially for three-dimensional systems.

\subsection{Self-diffusion}
\begin{figure}[htb]
\includegraphics[width=\columnwidth]{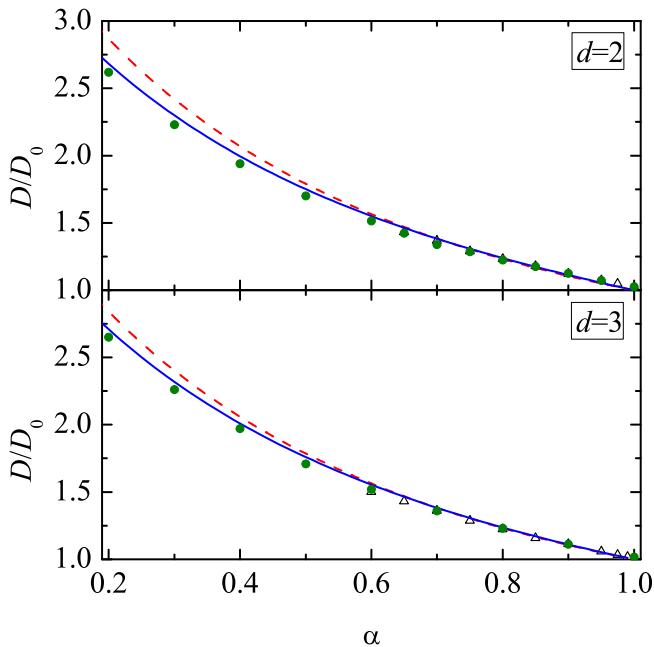}
\caption{(Color online) Plot of the  (reduced) self-diffusion
coefficient $D/D_0$ as a function of $\alpha$ for hard disks (top
panel) and hard spheres (bottom panel). The dashed and solid lines
represent the standard and modified first Sonine approximations,
respectively. The symbols are computer simulation results obtained
from the mean square displacement. In the top panel, the triangles
correspond to molecular dynamics results \protect\cite{BRMCGR00} and
the circles are DSMC results obtained in the present work. In the
bottom panel, the triangles are DSMC results reported in Ref.\
\protect\cite{BRMCGR00} and the circles are DSMC results presented
in Ref.\ \protect\cite{GM04} ($\alpha\geq 0.6$) and obtained in the
present work ($\alpha\leq 0.5$).
\label{fig6}}
\end{figure}

Finally, we consider the self-diffusion coefficient. This
coefficient has been measured in computer simulations from the mean
square displacement of a tagged particle in the HCS
\cite{BRMCGR00,LBD02,GM04}, as well as by the decay of a sinusoidal
perturbation in the concentration of tagged particles
\cite{BRMCGR00}. We observe in Fig.\ \ref{fig6} that both Sonine
approximations provide a general good agreement with simulation
data. However, the standard approximation slightly overestimates the
self-diffusion coefficient at high inelasticity, this effect being
corrected again by the modified approximation.

\section{Concluding remarks}
\label{sec6}
This work has been mainly motivated by the disagreement found at
high dissipation between the simulation data for the heat flux
transport coefficients and the expressions derived from the standard
first Sonine approximation \cite{BRM04,BRMMG05,MSG06}. Although this
disagreement appears beyond the region of inelasticity of practical
interest, it is physically   relevant from a fundamental point of
view to propose alternative theoretical approaches that correct the
limitations of the standard approximation. Here, we have
\kk{implemented} a modified version (\ref{2.30}) of the first Sonine
approximation (\ref{2.30x}), where the weight function is no longer
the Maxwell--Boltzmann distribution $f_M$ but the HCS distribution
$f^{(0)}$. Moreover, in order to preserve the solubility conditions,
the polynomial $\mathbf{S}(\mathbf{V})$ defined by Eq.\
(\ref{n4bis}) must be replaced by the modified polynomial
$\overline{\mathbf{S}}(\mathbf{V})$ defined by Eq.\ (\ref{2.31b}).
\kk{The idea behind the modified method is that the deviation of
$f^{(0)}$ from $f_M$ has an important influence on the NS
distribution $f^{(1)}$, so that the latter is better captured by the
approximation (\ref{2.30}) than by the approximation (\ref{2.30x}).
In other words, the rate of convergence of the polynomial expansion
is expected to be accelerated when $f^{(0)}$ rather than $f_M$ is
used as weight function.}

The structure of the transport coefficients is common in both
approximations. They are given by Eqs.\ (\ref{2.19})--(\ref{2.21})
and (\ref{2.29.3}). However, the $\alpha$-dependence of the
characteristic collision frequencies differs in both methods. In the
standard first Sonine approximation, those collision frequencies are
given by Eqs.\ (\ref{2.36x})--(\ref{2.37x}), while they are given by
Eqs.\ (\ref{2.36})--(\ref{2.37}) in the modified first Sonine
approximation. It is apparent that the distinction between both
approximations occurs in the value of the coefficient of the fourth
cumulant $a_2$. In the standard approximation that coefficient comes
from the dependence of the linearized Boltzmann collision operator
on $f^{(0)}$ only, while in the modified approximation it also comes
from the assumed form for $f^{(1)}$. The effect of the latter
contribution becomes more important than that of the former, so
that, for each collision frequency, the coefficient of $a_2$ changes
dramatically from the standard  approximation to the modified one.

As observed in Figs.\ \ref{fig1}--\ref{fig3}, the modified
approximation significantly improves the $\alpha$-dependence of
$\kappa$, $\mu$, and the difference $\kappa'=\kappa-n\mu/2T$. This
is the primary result of this paper. Additionally, as shown in
Figs.\ \ref{fig5} and \ref{fig6}, the slight discrepancies between
simulation and the standard first Sonine estimates for the shear
viscosity $\eta$ and the self-diffusion coefficient $D$ are
partially corrected by the modified approximation.

Although the results reported here have been restricted to a
low-density granular gas described by the inelastic Boltzmann
equation, they can be straightforwardly extended to finite density
in the framework of the Enskog kinetic theory. In that case,
application of the CE method shows that the effective collision
frequencies are the same as in the dilute limit, except for a
density-dependent factor $\chi$ \cite{GD99,L05}. The explicit
expressions for the NS transport coefficients are presented in
Appendix \ref{appB}. We expect that these Enskog results can
stimulate the performance of molecular dynamics simulations to test
whether or not the modified first Sonine approximation improves
again over the predictions of the standard approximation, especially
in the case of the heat flux transport coefficients at high
inelasticity.

\begin{acknowledgments}
This research  has been supported by the Ministerio de Educaci\'on y
Ciencia (Spain) through grants Nos.\ FIS2004-01399 (A.S. and V.G.)
and ESP2003-02859 (J.M.M.), partially financed by FEDER funds.
\end{acknowledgments}

\appendix
\section{Modified polynomial expansion\label{appC}}
\kk{Given a weight function
\beq
w(x)=e^{-x}\left[1+\sum_{k=2}^\infty a_k L_k^{(p-1)}(x)\right],
\label{C1}
\eeq
the mathematical problem consists of finding a set of polynomials
$\{\overline{L}_k^{(p)}(x)\}$ such that they are mutually orthogonal
with respect to the scalar product
\beq
(f_1,f_2)=\int_0^\infty dx\, x^p w(x) f_1(x)f_2(x).
\label{C2}
\eeq
}

 If $a_k=0$, then $w(x)=e^{-x}$ and one has the Laguerre
polynomials, i.e., $\overline{L}_k^{(p)}(x)={L}_k^{(p)}(x)$. In the
general case, the polynomials $\overline{L}_k^{(p)}(x)$ can be
obtained following the Gram--Schmidt orthogonalization procedure.
Suppose the polynomials $\overline{L}_\ell^{(p)}(x)$ with $\ell\leq
k-1$ are already known. The next unknown polynomial can be written
as
\beq
\overline{L}_{k}^{(p)}(x)=c_{k}^{(k)}x^{k}+\sum_{\ell=0}^{k-1}
c_\ell^{(k)} \overline{L}_{\ell}^{(p)}(x),
\label{C3}
\eeq
where the coefficients $c_\ell^{(k)}$ are to be determined. One of
them can be fixed by the standardization condition. For instance, we
can take $c_{k}^{(k)}=(-1)^k/k!$, which is the same coefficient as
in ${L}_{k}^{(p)}(x)$. The orthogonalization condition
$(\overline{L}_{\ell}^{(p)},\overline{L}_{k}^{(p)})=0$ for $\ell\leq
k-1$ gives
\beq
c_\ell^{(k)}=-c_k^{(k)}\frac{(\overline{L}_{\ell}^{(p)},x^k)}{(\overline{L}_{\ell}^{(p)},\overline{L}_{\ell}^{(p)})}.
\label{C4}
\eeq
This closes the construction of $\overline{L}_{k}^{(p)}(x)$ and the
process can be recursively continued. Since
$\overline{L}_{k-1}^{(p)}(x)x^{k+1}$ is a polynomial of degree $2k$,
and given the orthogonality properties of the Laguerre polynomials
$L_k^{p-1}(x)$ appearing in the representation (\ref{C1}), it is
straightforward to see that only the first $k$ coefficients
$\{a_\ell,\ell\leq k\}$ appear in $\overline{L}_{k}^{(p)}(x)$. The
norm of $\overline{L}_{k}^{(p)}(x)$, however, involves the
coefficient $a_{2k+1}$.

The three first polynomials are $\overline{L}_0^{(p)}(x)=1$,
\beq
\overline{L}_1^{(p)}(x)=(p+1)(1+a_2)-x,
\label{C5}
\eeq
\beq
\overline{L}_2^{(p)}(x)=\frac{1}{2}x^2+c_0^{(2)}+
c_1^{(2)}\overline{L}_1^{(p)}(x),
\label{C6}
\eeq
where
\beq
c_0^{(2)}=-\frac{p^2+3p+2}{2}(1+3a_2-a_3),
\label{C7}
\eeq
\begin{widetext}
\beq
c_1^{(2)}=\frac{p+2}{2}\frac{2+2(p+7)a_2-(p+1)a_2(3a_2-a_3)-(3p+11)a_3+(p+3)a_4}
{1+(p+4)a_2-(p+1)a_2^2-(p+2)a_3}.
\label{C8}
\eeq
\end{widetext}

\section{Transport coefficients for a dense granular gas\label{appB}}
In this Appendix, we give the expressions for  the NS transport
coefficients obtained from the Enskog kinetic equation by the
application of the CE method \cite{GD99,L05} in the first Sonine
approximation.

 The bulk
viscosity (which vanishes in the dilute limit) is
\begin{equation}
\label{B3}
\gamma=\eta_0\frac{2^{2d+1}}{(d+2)\pi}\phi^2 \chi
(1+\alpha)\left(1-\frac{1}{16} a_2\right),
\end{equation}
where
\beq
\phi\equiv \frac{\pi^{d/2}}{2^d\Gamma(1+d/2)} n\sigma^d
\label{B1}
\eeq
is the solid volume fraction and $\chi(\phi)$ is the pair
correlation function at contact. The shear viscosity $\eta$ has a
kinetic part $\eta^k$ and a collisional part $\eta^c$,  where
\begin{equation}
\label{B6}
\eta^k=\frac{\eta_0}{\chi}\left(\nu_{\eta}^*-\frac{1}{2}\zeta^*\right)^{-1}\left[1-\frac{2^{d-2}}{d+2}(1+\alpha)
(1-3 \alpha)\phi \chi \right],
\end{equation}
\begin{equation}
\label{B2}
\eta^c= \frac{2^{d-1}}{d+2}\phi \chi
(1+\alpha)\eta^k+\frac{d}{d+2}\gamma.
\end{equation}

Analogously, the coefficients associated with the heat flux have
also kinetic and collisional contributions. They are
\begin{eqnarray}
\label{B7}
\kappa^k&=&\frac{\kappa_0}{\chi}\frac{d-1}{d}\left(\nu_{\kappa}^*-2\zeta^*\right)^{-1}\Bigl\{1+2a_2
\nonumber\\
& &  +3\frac{
2^{d-3}}{d+2}\phi\chi(1+\alpha)^2\left[(1+\alpha)a_2-1+2\alpha\right]\Bigr\},\nn
&&
\end{eqnarray}
\beqa
\label{B4}
\kappa^c&=&3\frac{2^{d-2}}{d+2}\phi \chi
(1+\alpha)\kappa^k+\frac{2^{2d+1}(d-1)}{(d+2)^2\pi} \phi^2 \chi
(1+\alpha)\nn &&\times\left(1+\frac{7}{16} a_2 \right)\kappa_0,
\eeqa
\beqa
\mu^k&=&\frac{T}{n}\frac{\kappa_0}{\chi}\left(\nu_\kappa^*-\frac{3}{2}\zeta^*\right)^{-1}\left\{
\frac{\kappa^k}{\kappa_0}\xi\zeta^*+\frac{d-1}{d}a_2\right.\nn
&&\left.+3\frac{2^{d-3}(d-1)}{d(d+2)}\phi(\chi+\xi)(1+\alpha)\right.\nn
&&\left.\times\left[\frac{1}{6}(3\alpha^2-3\alpha+10+2d)a_2-\alpha(1-\alpha)\right]\right\},\nn
&&
\label{B8}
\eeqa
\begin{equation}
\label{B5}
\mu^c=\mu^k3\frac{2^{d-2}}{d+2}\phi \chi (1+\alpha).
\end{equation}
In Eq.\ (\ref{B8}), we have introduced the quantity $\xi\equiv
\partial(\phi\chi)/\partial\phi$.

Finally, the self-diffusion coefficient is simply given by
\begin{equation}
\label{B10}
D=\frac{D_0}{\chi}\frac{d+2}{2d}\frac{1}{\nu_D^*-\frac{1}{2}\zeta^*}.
\end{equation}

In the above equations, $\eta_0$, $\kappa_0$, and $D_0$ are the
elastic values of the low-density shear viscosity, thermal
conductivity, and self-diffusion coefficient, respectively. They are
given by
\begin{equation}
\label{2.22}
\eta_0=\frac{nT}{\nu_0},\quad
\kappa_0=\frac{d(d+2)}{2(d-1)}\frac{\eta_0}{m},
\end{equation}
\begin{equation}
\label{2.29.3bis}
D_0=\frac{2d}{d+2}\frac{T}{m\nu_0},
\end{equation}
 where the collision
frequency $\nu_0$ is defined by
\begin{equation}
\nu_0= \frac{8}{d+2}\frac{\pi^{(d-1)/2}}{\Gamma(d/2)}n\sigma
^{d-1}\left( \frac{T}{m} \right)^{1/2}.
\label{x3.14b}
\end{equation}
 In addition, the reduced cooling rate $\zeta^*$
for a dilute gas in the HCS is given by Eq.\ (\ref{x2.29}).

The expressions (\ref{B6})--(\ref{B10}) are common to the standard
and the modified first Sonine approximations. However, as discussed
in the main text, they differ in the $\alpha$-dependence of the
(reduced) effective collision frequencies $\nu_\eta^*$,
$\nu_\kappa^*$, and $\nu_D^*$, which are given by Eqs.\
(\ref{2.36x})--(\ref{2.37x}) in the standard approximation and by
Eqs.\ (\ref{2.36})--(\ref{2.37}) in the modified one. Note that,
since the collisional contributions  depend on their kinetic
counterparts, they also differ in both approximations.

\end{document}